\documentclass[seceq,preprint]{ptptex}

\newcommand{\dz}{\frac{dz}{2\pi i}}
\newcommand{\norm}[1]{{}^\times_\times{#1}^\times_\times}

\preprintnumber[3cm]{
hep-th/0303039\\YITP-03-8\\ March, 2003}

\title{
Hybrid Superstrings in NS-NS Plane Waves
}

\author{
Hiroshi \textsc{Kunitomo}%
}

\inst{
Yukawa Institute for Theoretical Physics\\
Kyoto University, Kyoto 606-8502, Japan
}

\recdate{March   , 2003}

\abst{
Using the hybrid formalism, superstrings in four-dimensional
NS-NS plane waves are studied in a manifest
supersymmetric manner. 
This description of the superstring is obtained through
a field redefinition of the RNS worldsheet fields
and defined as a topological $N=4$ string theory.
The Hilbert space consists of two types of representations
describing short and long strings. We study the physical spectrum 
to find boson-fermion asymmetry in the massless spectrum
of the short string. Some massive spectra of the short string 
and the massless spectrum of the long string are also studied.
}

\begin{document}

\maketitle

\section{Introduction}\label{intro}

Plane-wave backgrounds are exact string vacua and 
provide examples of string theories in non-compact 
curved space-times. These backgrounds are obtained 
by taking the Penrose limit of AdS spaces and have 
attracted much attention in the study of the AdS/CFT 
duality beyond the supergravity approximation.\cite{BMN} 
Despite much progress on this subject, the analyses performed
so far do not respect whole supersymmetry of these 
backgrounds. The superstring in R-R plane waves has 
been quantized in the light-cone gauge\cite{M} or using 
a non-covariant formalism.\cite{BM} Only a part of 
the supersymmetry is linearly realized in these formulations.
For superstrings in NS-NS plane waves, on the other hand,
we can use the Ramond-Neveu-Schwarz (RNS)
formalism for covariant quantization
in which, however, the space-time
supersymmetry is not manifest.
The manifest supersymmetry is not always necessary, 
but it is desirable to make transparent any cancellations 
coming from the supersymmetry. 

In this paper, we study four-dimensional superstrings 
in NS-NS plane waves in terms of the hybrid formalism
that was developed in Refs.~\citen{B} and \citen{BV} 
and has been applied to several compactified string 
theories.\cite{compact}
This description of the superstring is obtained through 
a field redefinition of the RNS worldsheet fields and 
manifestly preserves all isometries of the background, 
including supersymmetry.

Strings in NS-NS plane waves are described
by the Nappi-Witten (NW) model,\cite{NW} which is the WZW model 
on the group manifold $H_4$ generated by the four-dimensional 
Heisenberg algebra,
\begin{align}
  \left[{\cal J},{\cal P}\right]&={\cal P},\qquad
  \left[{\cal J},{\cal P}^*\right]=-{\cal P}^*,\nonumber\\
  \left[{\cal P},{\cal P}^*\right]&={\cal F}.\label{glbnw}
\end{align}
The Hilbert space of the NW model consists of two 
distinct representations, discrete (types II and III)
and continuous (type I). The model has a spectral 
flow symmetry and all flowed representations must also 
be included.\cite{KK,KP} 
The spectrally flowed discrete representations are 
regarded as describing short strings localized 
in the space transverse to the plane wave.
The flowed continuous representations define
long string states propagating 
in the whole four-dimensional space.
This structure is similar to the spectrum of 
the string in AdS$_3$,\cite{MO} 
which in fact is obtained by taking its Penrose 
limit.\cite{HS2}

The RNS superstring in this background is described
by superconformal field theories of the type
${H_4\times{\cal M}}$, where $H_4$ denotes the super NW
model and ${\cal M}$ is represented by an arbitrary $N=2$ 
unitary superconformal field theory 
with $c=9$.\cite{HS2}
This ${\cal M}$ sector represents a Calabi-Yau compactification 
if we project into the integral sector of the $U(1)_R$
charge $I_{{\cal M},0}$.
However, it was found in Ref.~\citen{HS2} that 
the string vacua have enhanced supersymmetry if we apply 
a generalized GSO projection that restricts the {\it total} 
$U(1)_R$ charge to integer values, while fractional $I_{{\cal M},0}$
is allowed. We adopt this weak GSO projection throughout. 

The hybrid superstrings in NS-NS plane waves are
related to the RNS superstrings by a field redefinition of
worldsheet fields. We show how to carry out
a redefinition from RNS fields to hybrid fields 
making all the space-time supersymmetry manifest. 
The model can be formulated as an $N=4$ topological string theory.\cite{BV} 

Next, we examine the physical spectrum at several lower mass levels.
We find that the massless spectrum of the short string
has boson-fermion asymmetry, which is allowed without breaking
supersymmetry. There are two massless bosons without fermionic
partners. This fact is made evident by the manifest 
supersymmetry. Some massive spectra of the short string 
and the massless spectrum of the long string 
are also studied in a manifestly supersymmetric manner.

This paper is organized as follows. In \S\ref{nsr}, 
we begin with a brief review of the super-NW model 
in the RNS formalism. The space-time supercharges 
are given in a form satisfying the conventional 
supersymmetry algebra.
Hybrid worldsheet fields are introduced
in \S\ref{hyb} through a redefinition of the RNS worldsheet 
fields. The model is then reformulated as a topological $N=4$ 
string theory.
In \S\ref{string}, we construct the Hilbert space 
of the hybrid superstring using hybrid fields. This 
space consists of two sectors representing short and long 
strings, respectively. The physical spectrum is studied 
in \S\ref{physspec}. It is found that the massless 
spectrum of the short string has boson-fermion asymmetry. 
The results are summarized with some discussion
in \S\ref{summary}.

\section{RNS superstrings in NS-NS plane waves}\label{nsr}

A description of RNS superstrings propagating in four-dimensional
NS-NS plane waves is provided by superconformal field theories 
of the type $H_4\times{\cal M}$.\cite{HS1,HS2} 
Here, $H_4$ denotes the super-NW model described by 
the super WZW model on the four-dimensional Heisenberg group. 
The Hilbert space of this model is constructed from 
representations of the $H_4$ super current algebra\footnote{
We consider only the holomorphic sector in this paper.
It can be easily combined with the anti-holomorphic sector
if necessary.\cite{HS2,KK,KP,MO}}
generated by bosonic $H_4$
currents, $(J,F,P,P^*)$, and their worldsheet superpartners,
$(\psi_J,\psi_F,\psi_P,\psi_{P^*})$:
\begin{alignat}{2}\label{h4sca}
 J(z)P(w)&\sim\frac{P(w)}{z-w},&\qquad
 J(z)P^*(w)&\sim-\frac{P^*(w)}{z-w},\nonumber\\
 P(z)P^*(w)&\sim\frac{1}{(z-w)^2}+\frac{F(w)}{z-w},&\qquad
 J(z)F(w)&\sim\frac{1}{(z-w)^2}, \nonumber\\
  \psi_P(z)\psi_{P^*}(w)&\sim\frac{1}{z-w},&\qquad
 \psi_J(z)\psi_F(w)&\sim\frac{1}{z-w},\nonumber\\
J(z)\psi_P(w)&\sim\psi_J(z)P(w)\sim 
\frac{\psi_P(w)}{z-w},&\qquad
J(z)\psi_{P^*}(w)&\sim\psi_J(z)P^*(w)\sim
-\frac{\psi_{P^*}(w)}{z-w},\nonumber\\
P(z)\psi_{P^*}(w)&\sim\psi_P(z)P^*(w)\sim
\frac{\psi_F(w)}{z-w}.& &
\end{alignat}
Representations of this algebra are easily obtained by using
a free-field realization,\cite{KK}
\begin{alignat}{2}
 J&=i\partial X^-,&\qquad F&=i\partial X^+,\nonumber\\
 P&=e^{iX^+}(i\partial Z+\psi^+\psi),&\qquad
 P^*&=e^{-iX^+}(i\partial Z^*-\psi^+\psi^*),\nonumber\\
\psi_F&=\psi^+,\qquad
\psi_J=\psi^-,&\qquad
\psi_P&=e^{iX^+}\psi,\qquad
\psi_{P^*}=e^{-iX^+}\psi^*,
\end{alignat}
where operator products of free fields are defined by
\begin{align}
 X^+(z)X^-(w)&\sim Z(z)Z^*(w)\sim -\log(z-w),\nonumber\\
\psi^+(z)\psi^-(w)&\sim\psi(z)\psi^*(w)\sim\frac{1}{z-w}.
\end{align}
The zero modes of bosonic currents provide generators of 
the space-time symmetry (\ref{glbnw}):
 \begin{alignat}{2}
 {\cal J}&=\oint\dz i\partial X^-,&\qquad
 {\cal F}&=\oint\dz i\partial X^+,\nonumber\\
 {\cal P}&=\oint\dz e^{iX^+}(i\partial Z+\psi^+\psi),&\qquad
 {\cal P}^*&=\oint\dz e^{-iX^+}(i\partial Z^*-\psi^+\psi^*).
\label{glbnwgen}
 \end{alignat}

The $N=1$ worldsheet superconformal symmetry is actually
enhanced to $N=2$, which is generated by
\begin{align}
 T_{H_4}&=-\partial X^+\partial X^--\partial Z\partial Z^*
 -\frac{1}{2}\psi^+\partial \psi^-
 -\frac{1}{2}\psi^-\partial\psi^+
 -\frac{1}{2}\psi\partial\psi^*
 -\frac{1}{2}\psi^*\partial\psi,\nonumber\\
 G^+_{H_4}&=\psi^+i\partial X^-+\psi i\partial Z^*,\qquad
G^-_{H_4}=\psi^-i\partial X^++\psi^*i\partial Z,\nonumber\\
 I_{H_4}&=\psi^+\psi^-+\psi\psi^*.
\end{align}
The model has central charge $c=6$, which is the same that for 
a superstring in flat four-dimensional space-time.

The ${\cal M}$ sector is represented by 
an arbitrary unitary representation of $N=2$ rational 
superconformal field theory with $c=9$. We denote the 
generators of this $N=2$ superconformal symmetry by 
$(T_{\cal M},G^\pm_{\cal M},I_{\cal M})$. 

In order to covariantly quantize the RNS superstring,
fermionic ghosts $(b,c)$ and bosonic ghosts $(\beta,\gamma)$
must be introduced. These superconformal ghosts satisfy 
\begin{equation}
 c(z)b(z)\sim\gamma(z)\beta(w)\ \sim\ \frac{1}{z-w},
\end{equation}
and have $N=1$ superconformal invariance generated by
\begin{align}
 T_{gh}&=-2b\partial c-\partial bc
-\frac{3}{2}\beta\partial\gamma-\frac{1}{2}\partial\beta\gamma,
\nonumber\\
 G_{gh}&=\frac{3}{2}\beta\partial c+\partial\beta c-2b\gamma.
\end{align}
The physical Hilbert space is defined by the cohomology 
${\cal H}_{{\rm phys}}={\rm Ker}Q_{{\rm BRST}}/{\rm Im}Q_{{\rm BRST}}$
of the BRST charge
\begin{eqnarray}
 Q_{{\rm BRST}}&=&\oint\dz\left(c\left(T_m+\frac{1}{2}T_{gh}\right)
+\gamma\left(G_m+\frac{1}{2}G_{gh}\right)\right),
\end{eqnarray}
where
\begin{eqnarray}
 T_m&=&T_{H_4}+T_{\cal M},\qquad 
G_m\ =\ G^+_{H_4}+G^-_{H_4}+G^+_{\cal M}+G^-_{\cal M}.
\end{eqnarray}

Then, we bosonize the worldsheet fermions and the $U(1)$
current in the ${\cal M}$ sector as
\begin{subequations}\label{bp}
\begin{align}
 \psi^+\psi^-&=i\partial H_0,\qquad
 \psi\psi^*=i\partial H_1,\\
 I_{\cal M}&=-\sqrt{3}i\partial H_2,\label{h2}
\end{align}
\end{subequations}
where the bosons $H_I(z)\ (I=0,1,2)$ satisfy
the standard OPEs:
\begin{equation}
 H_I(z)H_J(w)\sim -\delta_{IJ}\log(z-w).
\end{equation}
The superconformal ghosts are also bosonized by\cite{FMS}
\begin{align}
 c&=e^\sigma,\qquad b=e^{-\sigma},\nonumber\\
 \gamma&=\eta e^\phi=e^{\phi-\chi},\nonumber\\
 \beta&=e^{-\phi}\partial\xi=\partial\chi e^{-\phi+\chi},
\label{bg}
\end{align}
with
\begin{align}
 \phi(z)\phi(w)&\sim-\log(z-w),\nonumber\\
 \sigma(z)\sigma(w)&\sim\chi(z)\chi(w)\sim+\log(z-w).
\end{align}
Here, it is important that because the zero-mode $\xi_0$ is
not included in these formulas, the Hilbert space of 
the original bosonic ghosts $(\beta,\gamma)$ is different 
from that of the bosonized fields $(\phi,\xi,\eta)$ 
[or $(\phi,\chi)$]. The former (latter) is called the small 
(large) Hilbert space ${\cal H}_{{\rm small}}$ 
(${\cal H}_{{\rm large}}$). This extension of 
the Hilbert space is essential to realize
the supersymmetry.

In order to obtain a supersymmetric spectrum
in the RNS formalism, we must impose a GSO projection, which
guarantees the mutual locality of space-time supercharges.
If we use the weak GSO condition
\begin{eqnarray}
I_0&=&I_{H_4,0}+I_{{\cal M},0} \in \mathbb Z,\label{gso}
\end{eqnarray}
the model has the enhanced supersymmetry 
generated by the following four supercharges:\cite{HS2}
\begin{eqnarray}
 {\cal Q}^{\pm\pm}_{(-\frac{1}{2})}&=&\oint\dz
e^{-\frac{\phi}{2}}
e^{\pm iX^+}e^{\frac{i}{2}(H_0\pm(H_1+\sqrt{3}H_2))},\nonumber\\
 {\cal Q}^{\pm\mp}_{(-\frac{1}{2})}&=&\oint\dz
e^{-\frac{\phi}{2}}e^{\frac{i}{2}(-H_0\pm(H_1-\sqrt{3}H_2))}.
\label{susy}
\end{eqnarray}
The subscript ``$(-\frac{1}{2})$'' indicates that these operators
are given in the $-\frac{1}{2}$ picture, which is generally
read from the eigenvalue of the operator
\begin{equation}
 {\cal R}=\oint\dz(\xi\eta-\partial\phi).\label{picture}
\end{equation}

We note that these supercharges form a peculiar algebra,
due to the infinite degeneracy of pictures. This algebra is 
equivalent to supersymmetry only in the on-shell physical 
amplitudes. We change the picture for two of the supercharges, 
$({\cal Q}^{--}_{(-\frac{1}{2})},{\cal Q}^{+-}_{(-\frac{1}{2})})$, 
to $+\frac{1}{2}$ so that the conventional supersymmetry algebra
may hold without any condition:\footnote{
Rigorously speaking, the relative signs between terms in the explicit 
forms (\ref{susyp}) are not fixed without specifying the cocycle factors, 
which are usually omitted. This fact makes practical calculations difficult,
although it is often fixed by the Lorentz covariance of the result. This
complexity disappears in the hybrid formalism, and this is
actually one of the important advantages of the hybrid formalism.}
\begin{align}
 {\cal Q}^{--}_{(\frac{1}{2})}&=\oint\dz\left\{Q_{{\rm BRST}},\xi
e^{-\frac{\phi}{2}}e^{-iX^+}e^{\frac{i}{2}(H_0-(H_1+\sqrt{3}H_2))}\right\},
\nonumber\\
&=\oint\dz e^{-iX^+}\Bigg(
\eta be^{\frac{3}{2}\phi+\frac{i}{2}(H_0-(H_1+\sqrt{3}H_2))}
\nonumber\\
&\hspace{2.5cm}
+i\partial X^+e^{\frac{\phi}{2}-\frac{i}{2}(H_0+(H_1+\sqrt{3}H_2))}
+i\partial Z^*e^{\frac{\phi}{2}+\frac{i}{2}(-H_0-(H_1-\sqrt{3}H_2))}
\nonumber\\
&\hspace{2.5cm}
-\psi^+e^{\frac{\phi}{2}+\frac{i}{2}(H_0-(H_1+\sqrt{3}H_2))}
-G^-_{\cal M}e^{\frac{\phi}{2}+\frac{i}{2}(H_0-(H_1+\sqrt{3}H_2))}
\Bigg),\nonumber\\
 {\cal Q}^{+-}_{(\frac{1}{2})}&=\oint\dz\left\{Q_{{\rm BRST}},\xi 
e^{-\frac{\phi}{2}}e^{\frac{i}{2}(-H_0+(H_1-\sqrt{3}H_2))}\right\},
\nonumber\\
&=\oint\dz\Bigg(
\eta be^{\frac{3}{2}\phi+\frac{i}{2}(-H_0+(H_1-\sqrt{3}H_2))}
-i\partial X^-e^{\frac{\phi}{2}-\frac{i}{2}(-H_0-(H_1-\sqrt{3}H_2))}
\nonumber\\
&\hspace{2.5cm}
+i\partial Ze^{\frac{\phi}{2}-\frac{i}{2}(H_0+(H_1+\sqrt{3}H_2))}
-G^-_{\cal M}e^{\frac{\phi}{2}+\frac{i}{2}(-H_0+(H_1+\sqrt{3}H_2))}
\Bigg).\label{susyp}
\end{align}
Since we do not consider supercharges in other pictures,
we simply write $({\cal Q}^{\pm+}_{(-\frac{1}{2})},
{\cal Q}^{\pm-}_{(\frac{1}{2})})=({\cal Q}^{\pm+},{\cal Q}^{\pm-})$ 
in the rest of this paper. These picture-changed supercharges
together with (\ref{glbnwgen}) generate
the supersymmetry algebra for the NS-NS plane waves,\cite{HS2}
\begin{alignat}{2}
\left[{\cal J},{\cal P}\right]&={\cal P},&\qquad
\left[{\cal J},{\cal P}^*\right]&=-{\cal P}^*,\nonumber\\
\left[{\cal P},{\cal P}^*\right]&={\cal F},& &\nonumber\\
\left[{\cal J},{\cal Q}^{\pm\pm}\right]&=\pm{\cal Q}^{\pm\pm},
&\qquad
\left[{\cal J},{\cal Q}^{\pm\mp}\right]&=0,\nonumber\\
\left[{\cal Q}^{-+},{\cal P}\right]&=-{\cal Q}^{++},&\qquad
\left[{\cal Q}^{+-},{\cal P}^*\right]&={\cal Q}^{--},\nonumber\\
 \left\{{\cal Q}^{++},{\cal Q}^{--}\right\}&={\cal F},&\qquad
 \left\{{\cal Q}^{-+},{\cal Q}^{+-}\right\}&=-{\cal J},\nonumber\\
 \left\{{\cal Q}^{++},{\cal Q}^{+-}\right\}&={\cal P},&\qquad
 \left\{{\cal Q}^{-+},{\cal Q}^{--}\right\}&={\cal P}^*.\label{alg}
\end{alignat}

Before closing this section, it is useful to reconsider physical state
conditions in the RNS formalism. 
Although the physical states are defined by the BRST cohomology
in ${\cal H}_{{\rm small}}$, we must extend it to 
${\cal H}_{{\rm large}}$ to carry out a field redefinition to hybrid 
fields because, as mentioned above, 
${\cal H}_{{\rm small}}$ is not sufficient to realize the space-time
supersymmetry.
Therefore, we must generalize the physical state conditions to
\begin{subequations}\label{pscorg}
\begin{eqnarray}
 Q_{{\rm BRST}}|\psi\rangle&=&0,\nonumber\\
|\psi\rangle&\sim&|\psi\rangle+\delta|\psi\rangle,\qquad
\delta|\psi\rangle\ =\ Q_{{\rm BRST}}|\Lambda\rangle,\nonumber\\
\eta_0|\psi\rangle&=&\eta_0|\Lambda\rangle\ =\ 0,\label{psc}
\end{eqnarray}
where $|\psi\rangle, |\Lambda\rangle\in{\cal H}_{{\rm large}}$.
In addition to these cohomology conditions, 
we require that the physical states have ghost number one, i.e.,
\begin{equation}
  Q_{{\rm gh}}|\psi\rangle=|\psi\rangle,\label{gncond}
\end{equation}
\end{subequations}
counted by the charge\footnote{
This definition of the ghost number is related to 
the familiar one, $N_c=\oint\dz(cb-\gamma\beta)$, 
through the relation $Q_{gh}=N_c-{\cal R}$. 
The difference between the two definitions is a constant 
in a given picture.}
\begin{equation}
 Q_{{\rm gh}}=\oint\dz\left(cb-\xi\eta\right).
\end{equation}
The conditions given in (\ref{pscorg}) have a natural 
interpretation as a topological $N=4$ string theory.
We describe this interpretation below.

Let us note that there is the hidden twisted $N=4$ superconformal
symmetry generated by
\begin{align}
 T&=T_m+T_{{\rm gh}},\nonumber\\
 G^+&=J_{{\rm BRST}},\nonumber\\
 &= c\left(T_m+T_{\beta\gamma}\right)
+\gamma G_m-\gamma^2b+c\partial cb+\partial(c\xi\eta)+\partial^2c,
\nonumber\\
G^-&=b,\qquad
 \widetilde G^+\ =\ \eta,\qquad
 \widetilde G^-\ =\ \xi T-b\left\{Q_{{\rm BRST}},\xi\right\}+\partial^2\xi,
\nonumber\\
 I^{++}&=\eta c,\qquad
 I^{--}\ =\ b\xi,\qquad
 I\ =\ cb-\xi\eta,\label{topn4}
\end{align} 
in ${\cal H}_{{\rm large}}$. The physical state conditions (\ref{psc}) 
and the ghost number condition (\ref{gncond}) can be written 
in terms of these $N=4$ generators as
\begin{subequations}\label{psc1}
\begin{align}
 G^+_0|\psi\rangle&=0,\qquad 
\delta|\psi\rangle=G^+_0|\Lambda\rangle,\\
I_0|\psi\rangle&=|\psi\rangle,\\
\widetilde G^+_0|\psi\rangle&=\widetilde G^+_0|\Lambda\rangle=0,
\label{eta}
\end{align}
\end{subequations}
which can be naturally interpreted as physical state conditions
in the topological $N=4$ string theory.\cite{BV}
Because $\eta_0$-cohomology is trivial, we can always
solve Eq.~(\ref{eta}) by $|\psi\rangle=\widetilde G^+_0|V\rangle$ and
$|\Lambda\rangle=\widetilde G^+_0|\Lambda^-\rangle$ 
to rewrite (\ref{psc1}) in the more symmetric forms
\begin{subequations}\label{phys}
\begin{align}
 G^+_0\widetilde G^+_0|V\rangle&=0,\label{eom}\\
\delta|V\rangle&=G^+_0|\Lambda^-\rangle
+ \widetilde G^+_0|\widetilde\Lambda^-\rangle,\label{gauge}\\
I_0|V\rangle&=0.\label{u1}
\end{align}
\end{subequations}
In this paper, we call the condition (\ref{eom}) 
the equation of motion and the condition (\ref{gauge})
the gauge transformation, using the standard terminology in 
string field theory.
These conditions will be used to study the physical spectrum 
in \S\ref{physspec}.

\section{Hybrid superstrings in NS-NS plane waves}\label{hyb}

In this section, we develop the hybrid formalism 
for superstrings in NS-NS plane waves.
We first introduce hybrid fields by finding
a field redefinition from RNS fields, which allows the
whole space-time supersymmetry to be manifest.
We rewrite worldsheet superconformal generators using
these new fields to formulate the model as a topological
$N=4$ string theory.

As explained in the previous section,
the basic fields of the super NW model are free fields, 
$(X^\pm,Z,Z^*,\psi^\pm,\psi,\psi^*)$, and superconformal 
ghosts, $(b,c,\beta,\gamma)$. We add to them the boson $H_2$ 
coming from the $U(1)$ current (\ref{h2}) in the ${\cal M}$ sector,
which we need to define the supercharges (\ref{susy}).
Although the bosonic fields $(X^\pm,Z,Z^*)$ are common to 
the hybrid formalism,\footnote{
These bosons are not exactly the same in the two formalisms,
but they are related by the similarity 
transformation (\ref{similar}).} 
the remaining fields must be rearranged to obtain the
basic fields in the hybrid formalism.
Describing these fields in terms of the six free bosons 
$(H_0, H_1, H_2, \phi, \chi, \sigma)$ with the help of the
bosonization formulas (\ref{bp}) and (\ref{bg}), 
we carry out the linear transformation defined by
\begin{align}
 \phi_{--}&=-\frac{i}{2}H_0-\frac{i}{2}H_1
-\frac{i}{2}\sqrt{3}H_2+\frac{1}{2}\phi,
\nonumber\\
 \phi_{+-}&=\frac{i}{2}H_0+\frac{i}{2}H_1
-\frac{i}{2}\sqrt{3}H_2+\frac{1}{2}\phi,
\nonumber\\
 \phi_{++}&=-\frac{i}{2}H_0+\frac{i}{2}H_1
+\frac{i}{2}\sqrt{3}H_2-\frac{3}{2}\phi+\chi+\sigma,
\nonumber\\
 \phi_{-+}&=\frac{i}{2}H_0-\frac{i}{2}H_1
+\frac{i}{2}\sqrt{3}H_2-\frac{3}{2}\phi+\chi+\sigma,
\nonumber\\
 \rho&=\sqrt{3}H_2+3i\phi-2i\chi-i\sigma,
\nonumber\\
 \widehat H_2&=H_2+\sqrt{3}i\phi-\sqrt{3}i\chi,
\end{align}
and then define fermionic fields as
\begin{equation}
\theta^{\alpha\alpha'}=e^{\phi_{\alpha\alpha'}},\qquad
p_{\alpha\alpha'}=e^{-\phi_{\alpha\alpha'}},\qquad
(\alpha,\alpha'=\pm)
\label{bo}
\end{equation}
which satisfy
\begin{eqnarray}
 \theta^{\alpha\alpha'}(z)p_{\beta\beta'}(w)&\sim&
\frac{\delta^\alpha_\beta\delta^{\alpha'}_{\beta'}}{z-w}.
\end{eqnarray}
The basic fields of the hybrid superstrings are 
finally defined by Green-Schwarz-like fields with 
an additional boson: $(X^\pm,Z,Z^*,\theta^{\alpha\alpha'},
p_{\alpha\alpha'},\rho)$. 
The $U(1)$ boson in the ${\cal M}$ sector is also 
modified to $\widehat H_2$, which requires modifications of 
the superconformal generators to 
$(\widehat T_{\cal M}, \widehat G^\pm_{\cal M}, 
\widehat I_{\cal M})$, uniquely determined by the change of 
the $U(1)$ current
\begin{equation}
 \widehat I_{\cal M}=-\sqrt{3}i\partial\widehat H_2.
\end{equation}
We note that these new generators completely (anti-)commute 
with the hybrid fields.

The space-time supercharges (\ref{susy}) are written 
in terms of these hybrid fields: 
\begin{align}
 {\cal Q}^{++}&=\oint\dz e^{iX^+}p_{--},\nonumber\\
 {\cal Q}^{--}&=\oint\dz e^{-iX^+}\Big(
p_{++}+i\partial X^+\theta^{--}
+(i\partial Z^*+\theta^{-+}p_{++})\theta^{+-}
+e^{-i\rho}\theta^{-+}\widehat G^-_{\cal M}\Big),
\nonumber\\
 {\cal Q}^{-+}&=\oint\dz p_{+-},\nonumber\\
 {\cal Q}^{+-}&=\oint\dz\Big(
p_{-+}-i\partial X^-\theta^{+-}
+i\partial Z\theta^{--}
-e^{-i\rho}\theta^{++}\widehat G^-_{\cal M}\Big).\label{hsusy}
\end{align}
However, these supercharges are not symmetric.
This leads to a complicated hermiticity property
of hybrid fields.\cite{Bherm} These fields are
chiral coordinates in the sense that two of the supercharges,
${\cal Q}^{++}$ and ${\cal Q}^{-+}$, are simple 
superderivatives, $p_{--}$ and $p_{+-}$ (except for 
factors of $e^{\pm iX^+}$). 
In order to obtain symmetric supercharges and hybrid fields
with proper hermiticity, we must further carry out a similarity 
transformation generated by
\begin{align}
  U&=\oint\dz\Bigg(-e^{-i\rho}\theta^{++}\theta^{-+}\widehat G^-_M
+\frac{1}{2}i\partial X^+\theta^{++}\theta^{--}
+\frac{1}{2}i\partial Z^*\theta^{++}\theta^{+-}
\nonumber\\
&\hspace{2cm}
+\frac{1}{2}i\partial X^-\theta^{+-}\theta^{-+}
+\frac{1}{2}i\partial Z\theta^{-+}\theta^{--}
+\frac{1}{4}\theta^{-+}\theta^{++}\partial(\theta^{--}\theta^{+-})
\Bigg).\label{similar}
\end{align}
In fact, the space-time supercharges (\ref{hsusy})
have the symmetric forms
\begin{align}
{\cal Q}^{++}&=\oint\dz e^{iX^+}\Bigg(
p_{--}+\frac{1}{2}i\partial X^+\theta^{++}\nonumber\\
&\hspace{3cm}
+\frac{1}{2}(i\partial Z-\theta^{+-}p_{--})\theta^{-+}
+\frac{1}{8}\partial(\theta^{-+}\theta^{++})
\theta^{+-}\Bigg),\nonumber\\
{\cal Q}^{--}&=\oint\dz e^{-iX^+}\Bigg(
p_{++}+\frac{1}{2}i\partial X^+\theta^{--}\nonumber\\
&\hspace{3cm}
+\frac{1}{2}(i\partial Z^*+\theta^{-+}p_{++})\theta^{+-}
-\frac{1}{8}\partial(\theta^{--}\theta^{+-})\theta^{-+}\Bigg),
\nonumber\\
{\cal Q}^{-+}&=\oint\dz\left(
p_{+-}-\frac{1}{2}i\partial X^-\theta^{-+}+
\frac{1}{2}i\partial Z^*\theta^{++}
-\frac{1}{8}\partial(\theta^{-+}\theta^{++})\theta^{--}\right),
\nonumber\\
{\cal Q}^{+-}&=\oint\dz\left(
p_{-+}-\frac{1}{2}i\partial X^-\theta^{+-}+
\frac{1}{2}i\partial Z\theta^{--}
+\frac{1}{8}\partial(\theta^{--}\theta^{+-})\theta^{++}\right)
\label{susyh}
\end{align}
after the similarity transformation.

We can also rewrite the topological $N=4$ superconformal generators 
(\ref{topn4}) using the hybrid fields. 
The $N=2$ subalgebra is first given by 
\begin{align}
  T&=-\partial X^+\partial X^-
-\partial Z\partial Z^*
-p_{\alpha\alpha'}\partial\theta^{\alpha\alpha'}
+\frac{1}{2}\partial\rho\partial\rho
+\frac{1}{2}i\partial^2\rho
+\widehat T_{\cal M}
+\frac{1}{2}\partial \widehat I_{\cal M},\nonumber\\
G^+&=e^{-i\rho}\left(
d_{--}d_{+-}
+\frac{1}{8}\partial^2\theta^{-+}\theta^{++}
+\frac{1}{8}\theta^{-+}\partial^2\theta^{++}
-\frac{1}{4}\partial^2(\theta^{-+}\theta^{++})\right)
+\widehat G^+_M,\nonumber\\
G^-&=e^{i\rho}\left(
d_{-+}d_{++}+\frac{1}{8}\partial^2\theta^{--}\theta^{+-}
+\frac{1}{8}\theta^{--}\partial^2\theta^{+-}
-\frac{1}{4}\partial^2(\theta^{--}\theta^{+-})\right)
+\widehat G^-_M,\nonumber\\
I&=i\partial\rho-\sqrt{3}i\partial\widehat H_2,\label{uone}
\end{align}
where
\begin{align}
 d_{--}&=p_{--}-\frac{1}{2}i\partial X^+\theta^{++}
-\frac{1}{2}i\partial Z\theta^{-+}
+\frac{1}{4}\theta^{-+}\theta^{++}\partial\theta^{+-}
-\frac{1}{8}\partial(\theta^{-+}\theta^{++})\theta^{+-},
\nonumber\\
 d_{+-}&=p_{+-}+\frac{1}{2}i\partial X^-\theta^{-+}
-\frac{1}{2}i\partial Z^*\theta^{++}
-\frac{1}{4}\theta^{-+}\theta^{++}\partial\theta^{--}
+\frac{1}{8}\partial(\theta^{-+}\theta^{++})\theta^{--},
\nonumber\\
 d_{++}&=p_{++}-\frac{1}{2}i\partial X^+\theta^{--}
-\frac{1}{2}i\partial Z^*\theta^{+-}
-\frac{1}{4}\theta^{--}\theta^{+-}\partial\theta^{-+}
+\frac{1}{8}\partial(\theta^{--}\theta^{+-})\theta^{-+},
\nonumber\\
 d_{-+}&=p_{-+}+\frac{1}{2}i\partial X^-\theta^{+-}
-\frac{1}{2}i\partial Z\theta^{--}
+\frac{1}{4}\theta^{--}\theta^{+-}\partial\theta^{++}
-\frac{1}{8}\partial(\theta^{--}\theta^{+-})\theta^{++}
\label{scd}
\end{align}
are local currents of supercovariant derivatives.
It is also useful to introduce the bosonic supercovariant
derivatives as
\begin{align}
 \Pi^+&=i\partial X^++\frac{1}{2}\theta^{+-}\partial\theta^{-+}
-\frac{1}{2}\partial\theta^{+-}\theta^{-+},\nonumber\\
 \Pi^-&=i\partial X^--\frac{1}{2}\theta^{++}\partial\theta^{--}
+\frac{1}{2}\partial\theta^{++}\theta^{--},\nonumber\\
 \Pi_Z&=i\partial Z-\frac{1}{2}\theta^{++}\partial\theta^{+-}
+\frac{1}{2}\partial\theta^{++}\theta^{+-},\nonumber\\
 \Pi^*_Z&=i\partial Z^*-\frac{1}{2}\theta^{-+}\partial\theta^{--}
+\frac{1}{2}\partial\theta^{-+}\theta^{--}.\label{scmom}
\end{align}
These supercovariant derivatives and 
$\partial\theta^{\alpha\alpha'}$ form 
a closed superalgebra:
\begin{alignat}{2}
 d_{--}(z)d_{++}(w)&\sim
-\frac{\Pi^+(w)}{z-w},&\qquad
 d_{--}(z)d_{-+}(w)&\sim 
-\frac{\Pi_Z(w)}{z-w},\nonumber\\
d_{++}(z)d_{+-}(w)&\sim
-\frac{\Pi^*_Z(w)}{z-w},&\qquad
d_{+-}(z)d_{-+}(w)&\sim
\frac{\Pi^-(w)}{z-w},\nonumber\\
\Pi^+(z)\Pi^-(w)&\sim
\frac{1}{(z-w)^2},&\qquad
\Pi_Z(z)\Pi^*_Z(w)&\sim
\frac{1}{(z-w)^2},\nonumber\\
d_{--}(z)\Pi^-(w)&\sim
-\frac{\partial\theta^{++}(w)}{z-w},&\qquad
d_{--}(z)\Pi^*_Z(w)&\sim
-\frac{\partial\theta^{-+}(w)}{z-w},\nonumber\\
d_{++}(z)\Pi^-(w)&\sim
-\frac{\partial\theta^{--}(w)}{z-w},&\qquad
d_{++}(z)\Pi_Z(w)&\sim
-\frac{\partial\theta^{+-}(w)}{z-w},\nonumber\\
d_{+-}(z)\Pi^+(w)&\sim
\frac{\partial\theta^{-+}(w)}{z-w},&\qquad
d_{+-}(z)\Pi_Z(w)&\sim
-\frac{\partial\theta^{++}(w)}{z-w},\nonumber\\
d_{-+}(z)\Pi^+(w)&\sim
\frac{\partial\theta^{+-}(w)}{z-w},&\qquad
d_{-+}(z)\Pi^*_Z(w)&\sim
-\frac{\partial\theta^{--}(w)}{z-w}.
\end{alignat}
We note here that these supercovariant derivatives 
are {\it supercovariant} only in the sense that they 
(anti-)commute with two of the supercharges, ${\cal Q}^{\pm\mp}$.

It can be shown that the complicated forms of $G^\pm$ in (\ref{uone}) 
can be rewritten as
\begin{align}
  G^+&=e^{-i\rho}\norm{d_{--}d_{+-}}
+\widehat G^+_M,\nonumber\\
 G^-&=e^{i\rho}\norm{d_{-+}d_{++}}
+\widehat G^-_M
\end{align}
by introducing the new normal ordering ${}^\times_\times$ 
with respect to the currents $d_{\alpha\alpha'}$.
The generators (\ref{topn4}) of the topological 
$N=4$ superconformal symmetry are then given by
\begin{align}
  T&=-\partial X^+\partial X^-
-\partial Z\partial Z^*
-p_{\alpha\alpha'}\partial\theta^{\alpha\alpha'}
+\frac{1}{2}\partial\rho\partial\rho
+\frac{1}{2}i\partial^2\rho
+\widehat T_{\cal M}
+\frac{1}{2}\partial \widehat I_{\cal M},\nonumber\\
G^+&=e^{-i\rho}\norm{d_{--}d_{+-}}
+\widehat G^+_M,\nonumber\\
G^-&=e^{i\rho}\norm{d_{-+}d_{++}}
+\widehat G^-_M,\nonumber\\
\widetilde G^+&=
e^{2i\rho-\sqrt{3}i\widehat H_2}
\norm{d_{-+}d_{++}}
+e^{i\rho-\sqrt{3}i\widehat H_2}\widehat G^-_M,
\nonumber\\
\widetilde G^-&=
e^{-2i\rho+\sqrt{3}i\widehat H_2}
\norm{d_{--}d_{+-}}
+e^{-i\rho+\sqrt{3}i\widehat H_2}\widehat G^+_M,
\nonumber\\
I^{++}&=e^{i\rho-\sqrt{3}i\widehat H_2},\qquad
I^{--}=e^{-i\rho+\sqrt{3}i\widehat H_2},\nonumber\\
I&=i\partial\rho-\sqrt{3}i\partial\widehat H_2.\label{n4}
\end{align}
Physical states are defined by the conditions (\ref{phys}) 
using the zero modes $G_0^+$, $\widetilde G_0^+$ and $I_0$ of
these generators.
The supercharges (\ref{susyh}) (anti-)commute with them.
This guarantees the physical spectrum to be supersymmetric.

Finally, we rewrite the picture counting operator 
(\ref{picture}) interpreted in the hybrid formalism 
as the R-charge operator:
\begin{equation}
 {\cal R}=\oint\dz\left(
i\partial\rho-\frac{1}{2}(
\theta^{++}p_{++}+\theta^{-+}p_{-+}-
\theta^{--}p_{--}-\theta^{+-}p_{+-})\right).\label{Rch}
\end{equation}
This is useful to determine whether each component of the 
superfields is a space-time boson or fermion. The field with
(half-)integral R-charge is a space-time boson (fermion), because
it comes from the NS-(R-)sector in the RNS formalism.

\section{Spectral flow and the Hilbert space of 
the hybrid superstring}\label{string}

Now we study the Hilbert space of the hybrid superstring.
Using the hybrid fields, the $H_4$ currents are 
realized as
\begin{subequations}\label{hybfree}
\begin{alignat}{2}
 J&=i\partial X^-,&\qquad
 F&=i\partial X^+,\nonumber\\
 P&=e^{iX^+}(i\partial Z-\theta^{+-}p_{--}),&\qquad
 P^*&=e^{-iX^+}(i\partial Z^*+\theta^{-+}p_{++}).\label{nwhyb}
\end{alignat}
We can extend this $H_4$ current algebra to a superalgebra, 
which is an analog of the super current algebra (\ref{h4sca}), 
by introducing the space-time supercoordinates 
(and their conjugates)
\begin{alignat}{2}
 \Theta^{\pm\mp}&=\theta^{\pm\mp},&\qquad
 {\cal P}_{\pm\mp}&=p_{\pm\mp},\nonumber\\
 \Theta^{\pm\pm}&=e^{\pm iX^+}\theta^{\pm\pm},&\qquad 
{\cal P}_{\pm\pm}&=e^{\mp iX^+}p_{\pm\pm},\label{scoo}
\end{alignat}
\end{subequations}
together with an extra $U(1)$ current, $i\partial\rho$.
The Hilbert space of the hybrid superstring is constructed
from representations of this current superalgebra.
We can expand these currents as
\begin{alignat}{2}\label{currentmodes}
  J(z)&=\sum_nJ_nz^{-n-1},&\qquad
  F(z)&=\sum_nF_nz^{-n-1},\nonumber\\
  P(z)&=\sum_nP_nz^{-n-1},&\qquad
  P^*(z)&=\sum_nP^*_nz^{-n-1},\nonumber\\
  \Theta^{\pm\mp}(z)&=\sum_n\Theta^{\pm\mp}_nz^{-n},&\qquad
  {\cal P}_{\pm\mp}(z)&=\sum_n({\cal P}_{\pm\mp})_nz^{-n-1},\nonumber\\
  \Theta^{\pm\pm}(z)&=\sum_n\Theta^{\pm\pm}_nz^{-n},&\qquad
  {\cal P}_{\pm\pm}(z)&=\sum_n({\cal P}_{\pm\pm})_nz^{-n-1},
\nonumber\\
  i\partial\rho(z)&=\sum_n\rho_nz^{-n-1},&&
\end{alignat} 
where the mode operators satisfy the superalgebra
\begin{alignat}{2}
  \left[J_n,P_m\right]&=P_{n+m},&\qquad
  \left[J_n,P^*_m\right]&=-P_{n+m},\nonumber\\
  \left[J_n,F_m\right]&=n\delta_{n+m,0},&\qquad 
  \left[P_n,P^*_m\right]&=F_{n+m}+n\delta_{n+m,0},\nonumber\\
  \left[J_n,\Theta^{\pm\pm}_m\right]&=\pm\Theta^{\pm\pm}_{n+m},&\qquad
  \left[J_n,({\cal P}_{\pm\pm})_m\right]&=\mp({\cal
  P}_{\pm\pm})_{n+m},\nonumber\\
  \left[P_n,\Theta^{--}_m\right]&=-\Theta^{+-}_{n+m},&\qquad 
  \left[P_n,({\cal P}_{+-})_m\right]&=({\cal P}_{--})_{n+m},\nonumber\\
  \left[P^*_n,\Theta^{++}_m\right]&=\Theta^{-+}_{n+m},&\qquad
  \left[P^*_n,({\cal P}_{-+})_m\right]&=-({\cal P}_{++})_{n+m},\nonumber\\
  \left\{\Theta^{\pm\pm}_n,({\cal P}_{\pm\pm})_m\right\}&=\delta_{n+m,0},
&\qquad
  \left\{\Theta^{\pm\mp}_n,({\cal P}_{\pm\mp})_m\right\}&=\delta_{n+m,0},
\nonumber\\
\left[\rho_n,\rho_m\right]&=-n\delta_{n+m,0}.&&\label{hybcur}
\end{alignat}
Because the hybrid fields provide the free field 
realization (\ref{hybfree}), we can easily obtain the
representations of this superalgebra (\ref{hybcur}).
As in the $H_4$ (super) current algebra,\cite{KK,KP}$^,$\cite{HS2}
the only non-trivial point is the existence of the spectral flow
symmetry; i.e., the superalgebra (\ref{hybcur}) is 
preserved by the replacement
\begin{alignat}{2}
  J_n&\longrightarrow J_n,&\qquad
  F_n&\longrightarrow F_n+p\delta_{n,0},\nonumber\\
  P_n&\longrightarrow P_{n+p},&\qquad
  P^*_n&\longrightarrow P^*_{n-p},\nonumber\\
  \Theta^{\pm\mp}_n&\longrightarrow\Theta^{\pm\mp}_n,&\qquad
  ({\cal P}_{\pm\mp})_n&\longrightarrow ({\cal P}_{\pm\mp})_n,\nonumber\\
  \Theta^{\pm\pm}_n&\longrightarrow\Theta^{\pm\pm}_{n\pm p},&\qquad
  ({\cal P}_{\pm\pm})_n&\longrightarrow ({\cal P}_{\pm\pm})_{n\mp p},
\nonumber\\
 \rho_n&\longrightarrow\rho_n,& &
\label{spcf}
\end{alignat}
for any integer $p\in$\boldmath$\mathit{Z}$\unboldmath.
The Hilbert space contains all spectrally flowed representations 
classified into two types, describing short and long strings.\cite{KP}

\subsection{The Hilbert space of short strings}\label{short}

The Hilbert space of short strings in the hybrid formalism include
all spectrally flowed type II representations,\cite{KK} ($0<\eta<1$)
defined by
\begin{alignat}{2}
 J_0|j,\eta,p,l\rangle&=j|j,\eta,p,l\rangle,&\qquad
 F_0|j,\eta,p,l\rangle&=(\eta+p)|j,\eta,p,l\rangle,\nonumber\\
 J_n|j,\eta,p,l\rangle&=0,\quad (n>0)&\quad
 F_n|j,\eta,p,l\rangle&=0,\quad (n>0)\nonumber\\
 P_n|j,\eta,p,l\rangle&=0,\quad (n\ge-p)&\quad
 P^*_n|j,\eta,p,l\rangle&=0,\quad (n>p)\nonumber\\
 ({\cal P}_{\pm\mp})_n|j,\eta,p,l\rangle&=0,\quad (n\ge0)&\quad
 \Theta^{\pm\mp}_n|j,\eta,p,l\rangle&=0,\quad (n>0)\nonumber\\
 ({\cal P}_{++})_n|j,\eta,p,l\rangle&=0,\quad (n\ge p+1)&\quad
 \Theta^{++}_n|j,\eta,p,l\rangle&=0,\quad (n>-p-1)\nonumber\\
 ({\cal P}_{--})_n|j,\eta,p,l\rangle&=0,\quad (n\ge-p)&\quad
 \Theta^{--}_n|j,\eta,p,l\rangle&=0,\quad (n>p)\nonumber\\
 \rho_0|j,\eta,p,l\rangle&=(l-\eta)|j,\eta,p,l\rangle,&\quad
 \rho_n|j,\eta,p,l\rangle&=0,\quad (n>0)
\label{typeii}
\end{alignat}
where $l=0,\pm 1$. The $\rho_0$ eigenvalue is fixed so that 
the supercurrents $G^\pm$ in (\ref{n4}) are periodic
and a unique representative is selected from infinitely 
degenerate states whose existence is due to the pictures. 

The explicit representations are easily constructed in terms of 
the hybrid fields by noting that
the transverse fields $(Z,Z^*,\theta^{\pm\pm},p_{\pm\pm})$
satisfy the twisted boundary conditions:
\begin{alignat}{2}
   i\partial Z(e^{2\pi i}z)&=e^{-2\pi i\eta}i\partial Z(z),&\qquad
   i\partial Z^*(e^{2\pi i}z)&=e^{2\pi i\eta}i\partial Z^*(z),
\nonumber\\
   \theta^{\pm\pm}(e^{2\pi i}z)&=e^{\mp 2\pi i\eta}\theta^{\pm\pm}(z),
&\qquad
   p_{\pm\pm}(e^{2\pi i}z)&=e^{\pm 2\pi i\eta}p_{\pm\pm}(z).
\label{twstbc}
\end{alignat}
Then, the hybrid fields can be expanded as
\begin{alignat}{2}
   i\partial X^\pm(z)&=\sum_n\alpha^\pm_nz^{-n-1},&&\nonumber\\
  i\partial Z(z)&=\sum_nZ_{n+\eta}z^{-n-\eta-1},&\qquad
  i\partial Z^*(z)&=\sum_nZ^*_{n-\eta}z^{-n+\eta-1},
\nonumber\\
  \theta^{\pm\mp}(z)&=\sum_n\theta^{\pm\mp}_nz^{-n},&\qquad
  p_{\pm\mp}(z)&=\sum_n(p_{\pm\mp})_nz^{-n-1},\nonumber\\
  \theta^{\pm\pm}(z)&=\sum_n\theta^{\pm\pm}_{n\pm\eta}z^{-n\mp\eta},
&\qquad
  p_{\pm\pm}(z)&=\sum_n(p_{\pm\pm})_{n\mp\eta}z^{-n\pm\eta-1},
\label{ffr}
\end{alignat}
where the oscillators satisfy the canonical 
(anti-)commutation relations
\begin{alignat}{2}
  \left[\alpha^+_n,\alpha^-_m\right]&=n\delta_{n+m,0},&\qquad
  \left[Z_{n+\eta},Z^*_{m-\eta}\right]&=(n+\eta)\delta_{n+m,0},
\nonumber\\
  \left\{\theta^{\pm\mp}_n,(p_{\pm\mp})_m\right\}&=\delta_{n+m,0},&\qquad
  \left\{\theta^{\pm\pm}_{n\mp\eta},(p_{\pm\pm})_{m\pm\eta}\right\}&
=\delta_{n+m,0},
\label{oscccr}
\end{alignat}

The flowed type II representations are simply realized as 
Fock states of these oscillators (and $\rho_n$) on the ground state
 \begin{alignat}{2}  
 \alpha^-_0|\eta,\boldsymbol p,\boldsymbol\theta,l\rangle&=
j|\eta,\boldsymbol p,\boldsymbol\theta,l\rangle,&\qquad
 \alpha^+_0|\eta,\boldsymbol p,\boldsymbol\theta,l\rangle&=
(\eta+p)|\eta,\boldsymbol p,\boldsymbol\theta,l\rangle,\nonumber\\
 \alpha^\pm_n|\eta,\boldsymbol p,\boldsymbol\theta,l\rangle&=0,\quad
  (n>0)&&
\nonumber\\
 Z_{n+\eta}|\eta,\boldsymbol p,\boldsymbol\theta,l\rangle&=0,\quad (n\ge 0)&\qquad
 Z^*_{n-\eta}|\eta,\boldsymbol p,\boldsymbol\theta,l\rangle&=0,\quad
  (n>0)
\nonumber\\
 (p_{+-})_0|\eta,\boldsymbol p,\boldsymbol\theta,l\rangle&=
|\eta,\boldsymbol p,\boldsymbol\theta,l\rangle\frac{\partial}{\partial\theta},&\qquad
 \theta^{+-}_0|\eta,\boldsymbol p,\boldsymbol\theta,l\rangle&=
|\eta,\boldsymbol p,\boldsymbol\theta,l\rangle\theta,\nonumber\\
 (p_{-+})_0|\eta,\boldsymbol p,\boldsymbol\theta,l\rangle&=|\eta,\boldsymbol
  p,\boldsymbol\theta,l\rangle\frac{\partial}{\partial\bar\theta},
&\qquad
 \theta^{-+}_0|\eta,\boldsymbol p,\boldsymbol\theta,l\rangle&=
|\eta,\boldsymbol p,\boldsymbol\theta,l\rangle\bar\theta,\nonumber\\
 (p_{\pm\mp})_n|\eta,\boldsymbol p,\boldsymbol\theta,l\rangle&=0,\quad (n> 0)&\qquad
 \theta^{\pm\mp}_n|\eta,\boldsymbol
  p,\boldsymbol\theta,l\rangle&=0,\quad (n>0)
\nonumber\\
 (p_{++})_{n-\eta}|\eta,\boldsymbol p,\boldsymbol\theta,l\rangle&=0,
\quad (n>0)
&\qquad
 \theta^{++}_{n+\eta}|\eta,\boldsymbol p,\boldsymbol\theta,l\rangle&=0,
\quad (n\ge 0)\nonumber\\
 (p_{--})_{n+\eta}|\eta,\boldsymbol p,\boldsymbol\theta,l\rangle&=0,
\quad (n\ge 0)&\qquad
 \theta^{--}_{n-\eta}|\eta,\boldsymbol
  p,\boldsymbol\theta,l\rangle&=0,\quad (n>0)\nonumber\\
 \rho_0|\eta,\boldsymbol p,\boldsymbol\theta,l\rangle&=
(l-\eta)|\eta,\boldsymbol p,\boldsymbol\theta,l\rangle,&\qquad
 \rho_n|\eta,\boldsymbol p,\boldsymbol\theta,l\rangle&=0,\quad (n\ge0)
\label{ffgs}
 \end{alignat}
where we have diagonalized zero modes
$\alpha^\pm_0\ (=p^\pm)$ and $\theta^{\pm\pm}_0$, 
and denoted their eigenvalues by $\boldsymbol p=(j,p)$ and 
$\boldsymbol\theta=(\theta,\bar\theta)$. 
The short string states are obtained by multiplying the Fock states
by a superfield $\Psi(\boldsymbol p,\boldsymbol\theta)$ on which
$\frac{\partial}{\partial\theta}$ and
$\frac{\partial}{\partial\bar\theta}$ act.
Because $Z$ and $Z^*$ do not have zero modes, the short string 
is localized and cannot reach infinity in the transverse 
space.

The total Hilbert space is obtained as the tensor product of 
this Fock space and unitary representations 
describing the ${\cal M}$ sector. Because an arbitrary
unitary representation of the $N=2$ superconformal field theory is 
characterized by dimension $\Delta$ and $U(1)$ charge $Q$, 
we can formally define a unitary representation by
 \begin{align}
  \widehat L_{0,{\cal M}}|\Delta,Q\rangle&=
\Delta|\Delta,Q\rangle,\nonumber\\
  \widehat I_{0,{\cal M}}|\Delta,Q\rangle&=
Q|\Delta,Q\rangle.
 \end{align}
The short string is represented by
Fock states on the ground state
\begin{equation}
 |\eta,\boldsymbol p,\boldsymbol\theta,l;\Delta,Q\rangle=
|\eta,\boldsymbol p,\boldsymbol\theta,l\rangle\otimes
|\Delta,Q\rangle.\label{shortground}
\end{equation}

For later use, we note that this ground state has the following
eigenvalues:
 \begin{align}
  L_0|\eta,\boldsymbol p,\boldsymbol\theta,l;\Delta,Q\rangle&=
\Big((\eta+p)j-\frac{1}{2}(l-\eta)(l-\eta+1)\nonumber\\
&\hspace{1.8cm}
+\Delta-\frac{1}{2}Q-\frac{1}{2}\eta(1-\eta)\Big)
|\eta,\boldsymbol p,\boldsymbol\theta,l;\Delta,Q\rangle,
\label{totene}\\
  I_0|\eta,\boldsymbol p,\boldsymbol\theta,l;\Delta,Q\rangle&=
(l-\eta+Q)|\eta,\boldsymbol p,\boldsymbol\theta,l;\Delta,Q\rangle,
\label{toti}\\
 {\cal R}|\eta,\boldsymbol p,\boldsymbol\theta,l;\Delta,Q\rangle
&=|\eta,\boldsymbol p,\boldsymbol\theta,l;\Delta,Q\rangle\left(
\left(l-\frac{1}{2}\right)+\frac{1}{2}\left(\theta\frac{\partial}{\partial\theta}
-\bar\theta\frac{\partial}{\partial\bar\theta}\right)\right),\label{t2r}
 \end{align}
where the constant terms in (\ref{totene}) and (\ref{t2r})
are easily derived by using the bosonization (\ref{bo}).
The $U(1)$ charge condition (\ref{u1}) together with 
the $I_0$ eigenvalue (\ref{toti}) requires that the charge $Q$ 
of the short string be fractional.

\subsection{The Hilbert space of long strings}\label{long}

The long-string Hilbert space is given
by spectrally flowed type I representations 
$(\eta=0)$.\cite{KK} Mode expansions are 
easily obtained by setting $\eta=0$ in the previous 
expressions, except for the transverse coordinates 
$(Z,Z^*,\theta^{\pm\pm},p_{\pm\pm})$ 
having additional zero-modes. 
The Fock vacuum is defined by
 \begin{alignat}{2}
   \alpha^-_0|\boldsymbol p,\boldsymbol q,\boldsymbol\theta,
\boldsymbol{\tilde\theta},l\rangle&=
j|\boldsymbol p,\boldsymbol q,\boldsymbol\theta,\boldsymbol{\tilde\theta}
,l\rangle,&\qquad
 \alpha^+_0|\boldsymbol p,\boldsymbol q,\boldsymbol\theta,
\boldsymbol{\tilde\theta},l\rangle&=p|\boldsymbol p,
\boldsymbol q,\boldsymbol\theta,\boldsymbol{\tilde\theta},l\rangle,
\nonumber\\
 \alpha^\pm_n|\boldsymbol p,\boldsymbol q,
\boldsymbol\theta,\boldsymbol{\tilde\theta}
,l\rangle&=0,\quad (n>0)&&\nonumber\\
 Z_0|\boldsymbol p,\boldsymbol q,
\boldsymbol\theta,\boldsymbol{\tilde\theta}
,l\rangle&=q|\boldsymbol p,\boldsymbol q,
\boldsymbol\theta,\boldsymbol{\tilde\theta},l\rangle,&\qquad
 Z^*_0|\boldsymbol p,\boldsymbol q,
\boldsymbol\theta,\boldsymbol{\tilde\theta}
,l\rangle&=q^*|\boldsymbol p,
\boldsymbol q,\boldsymbol\theta,\boldsymbol{\tilde\theta},l\rangle,
\nonumber\\
 Z_n|\boldsymbol p,\boldsymbol q,
\boldsymbol\theta,\boldsymbol{\tilde\theta}
,l\rangle&=0,\quad (n>0)&\qquad
 Z^*_n|\boldsymbol p,\boldsymbol q,
\boldsymbol\theta,\boldsymbol{\tilde\theta},l\rangle&=0,\quad (n>0)
\nonumber\\
 (p_{+-})_0|\boldsymbol p,\boldsymbol q,
\boldsymbol\theta,\boldsymbol{\tilde\theta},l\rangle&=
|\boldsymbol p,\boldsymbol q,
\boldsymbol\theta,\boldsymbol{\tilde\theta}
,l\rangle\frac{\partial}{\partial\theta},&\qquad
 \theta^{+-}_0|\boldsymbol p,\boldsymbol q,
\boldsymbol\theta,\boldsymbol{\tilde\theta}
,l\rangle&=|\boldsymbol p,\boldsymbol q,
\boldsymbol\theta,\boldsymbol{\tilde\theta},l\rangle\theta,
\nonumber\\
 (p_{-+})_0|\boldsymbol p,\boldsymbol q,
\boldsymbol\theta,\boldsymbol{\tilde\theta},l\rangle&=
|\boldsymbol p,\boldsymbol q,
\boldsymbol\theta,\boldsymbol{\tilde\theta}
,l\rangle\frac{\partial}{\partial\bar\theta},&\qquad
 \theta^{-+}_0|\boldsymbol p,\boldsymbol q,
\boldsymbol\theta,\boldsymbol{\tilde\theta}
,l\rangle&=|\boldsymbol p,\boldsymbol q,
\boldsymbol\theta,\boldsymbol{\tilde\theta},l\rangle\bar\theta,
\nonumber\\
 (p_{\pm\mp})_n|\boldsymbol p,\boldsymbol q,
\boldsymbol\theta,\boldsymbol{\tilde\theta}
,l\rangle&=0,\quad (n> 0)&\qquad
 \theta^{\pm\mp}_n|\boldsymbol p,\boldsymbol q,
\boldsymbol\theta,\boldsymbol{\tilde\theta},l\rangle&=0,\quad (n>0)
\nonumber\\
 (p_{--})_0|\boldsymbol p,\boldsymbol q,
\boldsymbol\theta,\boldsymbol{\tilde\theta},l\rangle&=
|\boldsymbol p,\boldsymbol q,
\boldsymbol\theta,\boldsymbol{\tilde\theta}
,l\rangle\frac{\partial}{\partial\tilde\theta},&\qquad
 \theta^{--}_0|\boldsymbol p,\boldsymbol q,
\boldsymbol\theta,\boldsymbol{\tilde\theta}
,l\rangle&=|\boldsymbol p,\boldsymbol q,
\boldsymbol\theta,\boldsymbol{\tilde\theta},l\rangle
\tilde\theta,\nonumber\\
 (p_{++})_0|\boldsymbol p,\boldsymbol q,
\boldsymbol\theta,\boldsymbol{\tilde\theta},l\rangle&=
|\boldsymbol p,\boldsymbol q,\boldsymbol\theta,
\boldsymbol{\tilde\theta},l\rangle\frac{\partial}
{\partial\bar{\tilde\theta}},&\qquad
 \theta^{++}_0|\boldsymbol p,\boldsymbol q,
\boldsymbol\theta,\boldsymbol{\tilde\theta}
,l\rangle&=|\boldsymbol p,\boldsymbol q,
\boldsymbol\theta,\boldsymbol{\tilde\theta},l\rangle
 \bar{\tilde\theta},\nonumber\\
 (p_{--})_n|\boldsymbol p,\boldsymbol q,
\boldsymbol\theta,\boldsymbol{\tilde\theta}
,l\rangle&=0,\quad (n>0)&\qquad
 \theta^{--}_n|\boldsymbol p,\boldsymbol q,
\boldsymbol\theta,\boldsymbol{\tilde\theta},l\rangle&=0,\quad (n>0)
\nonumber\\
 (p_{++})_n|\boldsymbol p,\boldsymbol q,
\boldsymbol\theta,\boldsymbol{\tilde\theta}
,l\rangle&=0,\quad (n>0)&\qquad
 \theta^{++}_n|\boldsymbol p,\boldsymbol q,
\boldsymbol\theta,\boldsymbol{\tilde\theta},l\rangle&=0,\quad (n>0)
\nonumber\\
 \rho_0|\boldsymbol p,\boldsymbol q,
\boldsymbol\theta,\boldsymbol{\tilde\theta}
,l\rangle&=l|\boldsymbol p,\boldsymbol q,
\boldsymbol\theta,\boldsymbol{\tilde\theta}
,l\rangle,&\qquad
\rho_n|\boldsymbol p,\boldsymbol q,
\boldsymbol\theta,\boldsymbol{\tilde\theta}
,l\rangle&=0,\quad (n>0)
 \end{alignat} 
where $\boldsymbol q=(q,q^*)$ and $\boldsymbol{\tilde\theta}=
(\tilde\theta,\bar{\tilde\theta})$ are the additional zero modes.
The coefficient superfield in this sector is a function of the
zero modes $(\boldsymbol p,\boldsymbol q,
\boldsymbol\theta,\boldsymbol{\tilde\theta})$.
The long strings can freely propagate in the four-dimensional
space  $(X^\pm,Z,Z^*)$.

The long string is represented by Fock states on the ground state
\begin{equation}
 |\boldsymbol p,\boldsymbol q,
\boldsymbol\theta,\boldsymbol{\tilde\theta},l;\Delta,Q\rangle=
 |\boldsymbol p,\boldsymbol q,
\boldsymbol\theta,\boldsymbol{\tilde\theta},l\rangle
\otimes|\Delta,Q\rangle,
\end{equation}
having the following eigenvalues:
\begin{align} 
 L_0|\boldsymbol p,\boldsymbol q,
\boldsymbol\theta,\boldsymbol{\tilde\theta},l;\Delta,Q\rangle&=
 (pj+qq^*-\frac{1}{2}l(l+1)+\Delta-\frac{1}{2}Q)
|\boldsymbol p,\boldsymbol q,
\boldsymbol\theta,\boldsymbol{\tilde\theta},l;\Delta,Q\rangle,\\
 I_0|\boldsymbol p,\boldsymbol q,
\boldsymbol\theta,\boldsymbol{\tilde\theta},l;\Delta,Q\rangle&=
 (l+Q)|\boldsymbol p,\boldsymbol q,
\boldsymbol\theta,\boldsymbol{\tilde\theta},l;\Delta,Q\rangle,
\label{contu1}\\
 {\cal R}|\boldsymbol p,\boldsymbol q,
\boldsymbol\theta,\boldsymbol{\tilde\theta},l;\Delta,Q\rangle&=
 |\boldsymbol p,\boldsymbol q,
\boldsymbol\theta,\boldsymbol{\tilde\theta},l;\Delta,Q\rangle
 \left(l+\frac{1}{2}\left(\theta\frac{\partial}{\partial\theta}
 +\tilde\theta\frac{\partial}{\partial\tilde\theta}
-\bar\theta\frac{\partial}{\partial\bar\theta}
-\bar{\tilde\theta}\frac{\partial}{\partial\bar{\tilde\theta}}\right)\right).
\label{utotlong}
\end{align}
The $U(1)$ charge condition (\ref{u1}) together with 
the $I_0$ eigenvalue (\ref{contu1}) leads to the result 
that the long string must have integral $Q$.

\section{Physical spectrum}\label{physspec}

In this section
we study the physical spectrum at lower mass levels explicitly.
We concentrate on the states whose ${\cal M}$ sector is 
composed of (anti-)chiral primary states characterized 
by $\Delta=\frac{|Q|}{2}$. Then we solve the physical state 
conditions (\ref{phys}).

\subsection{Physical states in the short string sector}

We first examine physical states at mass levels 
$N=0,\eta,1-\eta$ in the short string sector.
It can easily be shown that it is sufficient to study the $l=1,0$ 
cases, since there is no physical state with 
$\rho_0$-momentum $l=-1$ at these levels.
The $U(1)$ charge condition (\ref{u1}) and the chirality 
condition $\Delta=\frac{|Q|}{2}$ yield
$\Delta=-\frac{Q}{2}=\frac{1}{2}(1-\eta)$ for the $l=1$ case
and $\Delta=\frac{Q}{2}=\frac{\eta}{2}$ for the $l=0$ case. 

Let us start by considering the oscillator ground state $N=0$ 
with $l=1$, given by
\begin{equation}
|V\rangle=|1\rangle\Psi^{(\frac{1}{2})}
(\boldsymbol p,\boldsymbol\theta).
\end{equation}
We denote here the state (\ref{shortground}) with $l=1$ by 
$|1\rangle$ and use this abbreviation in this subsection
for simplicity. A half of the supersymmetry is realized on 
the coefficient superfield $\Psi^{(\frac{1}{2})}$ by\footnote{
The other half is a part of the DDF operators discussed in
\S\ref{summary}. These operators relate the states 
with different masses.}
 \begin{equation}
  Q^{-+}=\frac{\partial}{\partial\theta}-\frac{1}{2}j\bar\theta,\qquad
  Q^{+-}=\frac{\partial}{\partial\bar\theta}-\frac{1}{2}j\theta.
\label{susyforcomp}
 \end{equation}
The superscript ``$(\frac{1}{2})$'' of the coefficient superfield indicates 
that its first component has R-charge $\frac{1}{2}$, 
which can be read from Eq.~(\ref{t2r}). 
The physical state conditions (\ref{phys}) 
lead to manifestly supersymmetric conditions on the superfield:
\begin{subequations}\label{tac}
 \begin{eqnarray}
  D\bar D\Psi^{(\frac{1}{2})}&=&0,\label{taceom}\\
  \delta\Psi^{(\frac{1}{2})}&=&\bar D\Lambda^{(1)},\label{tacgau}
 \end{eqnarray}
\end{subequations}
where $\Lambda^{(1)}$ is an arbitrary gauge parameter superfield
and supercovariant derivatives are defined by
 \begin{equation}
  D=\frac{\partial}{\partial\theta}+\frac{1}{2}j\bar\theta,\qquad
 \bar D=\frac{\partial}{\partial\bar\theta}+\frac{1}{2}j\theta.
\label{supercov}
 \end{equation}

The conditions (\ref{tac}) can be easily solved by choosing 
an appropriate gauge as 
\begin{equation}
 \Psi^{(\frac{1}{2})}=\bar\theta\bar\phi^{(0)}(p,j=0).\label{tachyon}
\end{equation}
The physical component $\bar\phi^{(0)}$ is a space-time boson 
and is identified with the {\it tachyon-like} state obtained 
in Ref.~\citen{HS2}. The solution (\ref{tachyon}) also shows 
that there is no fermionic massless physical state; 
i.e., the physical spectrum has boson-fermion asymmetry. 
This is only possible for the massless $(j=0)$ state,
on which the supercharges (\ref{susyforcomp}) anti-commute.

For the oscillator ground state with $l=0$,
\begin{equation}
  |V\rangle=
|0\rangle\Psi^{(-\frac{1}{2})}(\boldsymbol p,\boldsymbol\theta),
\end{equation}
physical state conditions are provided by
 \begin{eqnarray}
  \bar D D\Psi^{(-\frac{1}{2})}&=&0,\nonumber\\
  \delta\Psi^{(-\frac{1}{2})}&=&D\Lambda^{(-1)},
 \end{eqnarray}
and the solution has a form similar to $(\ref{tachyon})$:
\begin{equation}
 \Psi^{(-\frac{1}{2})}=\theta\phi^{(0)}(p,j=0).\label{graviton}
\end{equation}
The massless boson $\phi^{(0)}$ has no fermionic partner
and is identified with the {\it graviton-like} state obtained
in Ref.~\citen{HS2}.

Next, we consider two massive cases, $N=\eta$ and $1-\eta$.
General states at the level $N=\eta$ are expanded
in three Fock states as
\begin{equation}
 |V\rangle=
(\Pi^*_Z)_{-\eta}|l\rangle\Psi^{(l-\frac{1}{2})}(\boldsymbol p,\boldsymbol\theta)
+(d_{++})_{-\eta}|l\rangle\Phi^{(l)}(\boldsymbol p,\boldsymbol\theta)
+\theta^{--}_{-\eta}|l\rangle\Xi^{(l)}(\boldsymbol p,\boldsymbol\theta).
\end{equation}
Because we use a supercovariant basis created by 
$((\Pi^*_Z)_{-\eta},(d_{++})_{-\eta},\theta^{--}_{-\eta})$, 
the coefficient fields are superfields; i.e. their 
supersymmetry transformations are generated by the
supercharges (\ref{susyforcomp}).
The equations of motion for the $l=1$ case can be written
 \begin{align}
  D\left(\bar D\Xi^{(1)}+\eta\Psi^{(\frac{1}{2})}\right)&=0,
\nonumber\\
  \bar D\left(\Xi^{(1)}-(p+\eta)D\Psi^{(\frac{1}{2})}\right)+
\left((p+\eta)j+\eta\right)\Psi^{(\frac{1}{2})}&=0,
 \end{align}
with the gauge transformations
 \begin{align}
  \delta\Psi^{(\frac{1}{2})}&=\bar D\Lambda^{(1)},\nonumber\\
  \delta\Phi^{(1)}&=\Sigma^{(1)},\nonumber\\
  \delta\Xi^{(1)}&=-\eta\Lambda^{(1)}.
\end{align}
Choosing the $\Phi^{(1)}=\Xi^{(1)}=0$ gauge, 
the physical state is described by an anti-chiral superfield 
satisfying $D\Psi^{(\frac{1}{2})}=0$ and the on-shell condition
$((p+\eta)j+\eta)\Psi^{(\frac{1}{2})}=0$.
The anti-chiral superfield has the explicit form
\begin{equation}
 \Psi^{(\frac{1}{2})}=\psi^{(\frac{1}{2})}
+\bar\theta\bar\phi^{(0)}
-\frac{1}{2}\theta\bar\theta j\psi^{(\frac{1}{2})},
\end{equation}
containing one boson, $\bar\phi^{(0)}$, and 
one fermion, $\psi^{(\frac{1}{2})}$.

For the case $l=0$, the equations of motion
\begin{align}
 \bar D\left(\Xi^{(0)}-(p+\eta)D\Psi^{(-\frac{1}{2})}\right)&=0,
\nonumber\\
D\left(\Xi^{(0)}-(p+\eta)\Phi^{(0)}\right)&=0,\nonumber\\
(p+\eta)\bar DD\Phi^{(0)}+\eta D\Psi^{(-\frac{1}{2})}
+\left[D,\bar D\right]\Xi^{(0)}&=\left((p+\eta)j+\eta\right)\Phi^{(0)},
\nonumber\\
(p+\eta)D\left(\bar D\Xi^{(0)}+\eta\Psi^{(-\frac{1}{2})}\right)&=
\left((p+\eta)j+\eta\right)\Xi^{(0)}
\end{align}
and the gauge transformations
\begin{align}
  \delta\Psi^{(-\frac{1}{2})}&=\Sigma^{(-\frac{1}{2})}
+D\Lambda^{(-1)},\nonumber\\
\delta\Phi^{(0)}&=D\Sigma^{(-\frac{1}{2})},\nonumber\\
\delta\Xi^{(0)}&=(p+\eta)D\Sigma^{(-\frac{1}{2})}
\end{align}
can be solved by choosing $\Psi^{(-\frac{1}{2})}=0$ gauge as
\begin{equation}
 \Xi^{(0)}=(p+\eta)\frac{1}{j}\bar DD\Phi^{(0)}.
\end{equation}
The physical state is identified with an unconstrained 
superfield $\Phi^{(0)}$ satisfying the on-shell condition
$\left((p+\eta)j+\eta\right)\Phi^{(0)}=0$. 

We can easily repeat the above analysis to study the physical spectrum 
at the level $N=1-\eta$. The states at this level are generally
\begin{equation}
 |V\rangle=(\Pi_Z)_{-1+\eta}|l\rangle
\Psi^{(l-\frac{1}{2})}(\boldsymbol p,\boldsymbol\theta)
+(d_{--})_{-1+\eta}|l\rangle\Phi^{(l-1)}(\boldsymbol p,\boldsymbol\theta)
+\theta^{++}_{-1+\eta}|l\rangle\Xi^{(l-1)}(\boldsymbol p,\boldsymbol\theta).
\end{equation}
For $l=1$, the equations of motion and the gauge
transformations are given by
 \begin{align}
  D\left(\Xi^{(0)}-(p+\eta)\bar D\Psi^{(\frac{1}{2})}\right)&=0,
\nonumber\\
  (p+\eta)D\bar D\Phi^{(0)}
+(1-\eta)\bar D\Psi^{(\frac{1}{2})}
-\left[D,\bar D\right]\Xi^{(0)}&=
\left((p+\eta)j+1-\eta\right)\Phi^{(0)},\nonumber\\
  (p+\eta)\bar D\left(D\Xi^{(0)}+(1-\eta)\Psi^{(\frac{1}{2})}\right)
&=\left((p+\eta)j+1-\eta\right)\Xi^{(0)},\nonumber\\
\bar D\left(\Xi^{(0)}-(p+\eta)\Phi^{(0)}\right)&=0,
 \end{align}
and
 \begin{align}
  \delta\Psi^{(\frac{1}{2})}&=
     \Sigma^{(\frac{1}{2})}+\bar D\Lambda^{(1)},\nonumber\\
  \delta\Phi^{(0)}&=
     \bar D\Sigma^{(\frac{1}{2})},\nonumber\\
  \delta\Xi^{(0)}&=
     (p+\eta)\bar D\Sigma^{(\frac{1}{2})},
 \end{align}
respectively. The physical state is given by an unconstrained
superfield $\Phi^{(0)}$ satisfying
$\left((p+\eta)j+1-\eta\right)\Phi^{(0)}=0$.
The superfield $\Psi^{(\frac{1}{2})}$ can be gauged away,
and $\Xi^{(0)}$ is expressed in terms of $\Phi^{(0)}$.

We can also solve the physical state conditions for $l=0$,
 \begin{align}
  D\left(\Xi^{(-1)}-(p+\eta)\bar D\Psi^{(-\frac{1}{2})}\right)
   +\left((p+\eta)j+1-\eta\right)\Psi^{(-\frac{1}{2})}&=0,\nonumber\\
  \bar D\left(D\Xi^{(-1)}+(1-\eta)\Psi^{(-\frac{1}{2})}\right)&=0,
\end{align}
and
\begin{align}
  \delta\Psi^{(-\frac{1}{2})}&=D\Lambda^{(-1)},\nonumber\\
  \delta\Phi^{(-1)}&=\Sigma^{(-1)},\nonumber\\
  \delta\Xi^{(-1)}&=-(1-\eta)\Lambda^{(-1)},
\end{align}
in terms of a chiral superfield $\Psi^{(-\frac{1}{2})}$ satisfying
\begin{align}
  \bar D\Psi^{(-\frac{1}{2})}&=0,\nonumber\\
  \left((p+\eta)j+1-\eta\right)\Psi^{(-\frac{1}{2})}&=0. 
\end{align}
The explicit form of this chiral superfield is 
\begin{equation}
 \Psi^{(-\frac{1}{2})}=\psi^{(-\frac{1}{2})}+\theta\phi^{(0)}+
\frac{1}{2}\theta\bar\theta j\psi^{(-\frac{1}{2})}.
\end{equation}

In short, the physical spectrum at these massive levels
contains two types of multiplets, (anti-)chiral and
unconstrained. The latter is reducible and
decomposes into two (chiral and anti-chiral) multiplets.

\subsection{Physical states in the long string sector}

In the long string sector, we examine only the ground state 
with $\Delta=Q=0$. The $U(1)$ charge condition (\ref{u1}) 
leads to $l=0$, and therefore the state is given by
\begin{equation}
 |V\rangle=|\boldsymbol p,\boldsymbol q,
\boldsymbol\theta,\boldsymbol{\tilde\theta},0;0,0\rangle 
V^{(0)}(\boldsymbol p,\boldsymbol q,
\boldsymbol\theta,\boldsymbol{\tilde\theta}).
\end{equation}
The physical state conditions become
\begin{subequations}\label{longcondition}
 \begin{align}
 & \left(D\bar D\bar{\tilde D}\tilde D-
\tilde D\bar D\bar{\tilde D}D\right)V^{(0)}=0,\label{max}\\
 & \delta V^{(0)}=\tilde DD\Lambda^{(-1)}+\bar D\bar{\tilde D}\bar\Lambda^{(1)},
 \end{align}
\end{subequations}
where the supercovariant derivatives in this sector have the forms
\begin{alignat}{2}
 D&=\frac{\partial}{\partial\theta}+\frac{1}{2}j\bar\theta
-\frac{1}{2}q^*\bar{\tilde\theta},&\qquad
 \bar D&=\frac{\partial}{\partial\bar\theta}+\frac{1}{2}j\theta
-\frac{1}{2}q\tilde\theta,\nonumber\\
 \tilde D&=\frac{\partial}{\partial\tilde\theta}
-\frac{1}{2}p\bar{\tilde\theta}
-\frac{1}{2}q\bar\theta,&\qquad
 \bar{\tilde D}&=\frac{\partial}{\partial\bar{\tilde\theta}}
-\frac{1}{2}p\tilde\theta
-\frac{1}{2}q^*\theta.
\end{alignat}
The conditions (\ref{longcondition}) are essentially
the same as those for the four-dimensional vector
multiplet. In the WZ gauge,
 \begin{align}
  V=&\frac{1}{2}\tilde\theta\bar{\tilde\theta}A_+
   -\frac{1}{2}\theta\bar\theta A_-
   +\frac{1}{2}\tilde\theta\bar\theta A
   +\frac{1}{2}\theta\bar{\tilde\theta}A^*\nonumber\\
  &
   +\theta\bar\theta\bar{\tilde\theta}\tilde\lambda
   -\tilde\theta\bar\theta\bar{\tilde\theta}\lambda
   +\theta\tilde\theta\bar\theta\bar{\tilde\lambda}
   -\theta\tilde\theta\bar{\tilde\theta}\bar\lambda
   +\theta\tilde\theta\bar\theta\bar{\tilde\theta}{\cal D},
 \end{align}
the equations of motion (\ref{max}) lead to 
the Maxwell equations
 \begin{align}
  &\frac{1}{4}j(pA_-+jA_++qA^*+q^*A)
-\frac{1}{2}(pj+qq^*)A_-=0,\nonumber\\
  &\frac{1}{4}p(pA_-+jA_++qA^*+q^*A)
-\frac{1}{2}(pj+qq^*)A_+=0,\nonumber\\
  &\frac{1}{4}q^*(pA_-+jA_++qA^*+q^*A)
-\frac{1}{2}(pj+qq^*)A^*=0,\nonumber\\
  &\frac{1}{4}q(pA_-+jA_++qA^*+q^*A)
-\frac{1}{2}(pj+qq^*)A=0,
\end{align}
the massless Dirac equations
\begin{align}
(q^*\bar{\tilde\lambda}-j\bar\lambda)&=0,\qquad
(p\bar{\tilde\lambda}+q\bar\lambda)=0,\nonumber\\
(q\tilde\lambda-j\lambda)&=0,\qquad
(p\tilde\lambda+q^*\lambda)=0,
\end{align}
and ${\cal D}=0$ for the auxiliary field.
The massless spectrum of the long string is thus 
the vector multiplet in the four-dimensional 
{\it free-field space} $(X^\pm,Z,Z^*)$.

\section{Summary and Discussion}\label{summary}

In this paper, we have studied four-dimensional superstrings
in NS-NS plane-wave backgrounds using the hybrid 
formalism. This description of the superstring has been 
obtained through a field redefinition of the worldsheet 
fields in the super-NW model.\cite{HS2}
Because we have adopted a weak GSO projection that
restricts only the total $U(1)_R$ charge to integer values,
the model has enhanced supersymmetry, which is manifest 
in the hybrid formalism. The Hilbert space consists of 
two sectors, describing short and long strings,
and including all the spectrally flowed representations of 
types II and I, respectively.\cite{KK,KP} 
Then, we studied physical states to find boson-fermion 
asymmetry in the massless spectrum of the short string. 
There are two massless bosons, called tachyon-like 
and graviton-like in Ref.~\citen{HS2}, but no fermionic 
partners. We have also identified massive physical 
states at the levels $N=\eta$ and $1-\eta$ 
in the short string sector
and massless physical states in the long string sector. 
The massless physical spectrum of the long string is 
the vector multiplet freely propagating 
in the four-dimensional space $(X^\pm,Z,Z^*)$.

The massive physical states obtained by solving the physical
state conditions are also created by acting with the DDF operators
\begin{align}
   {\cal P}_n=&\oint\dz 
e^{i\left(\frac{n+\eta}{p+\eta}\right)X^+}
\left(i\partial Z-\left(\frac{n+\eta}{p+\eta}\right)
\theta^{+-}p_{--}\right),\nonumber\\
 {\cal P}^*_n=&\oint\dz 
e^{i\left(\frac{n-\eta}{p+\eta}\right)X^+}
\left(i\partial Z^*-\left(\frac{n-\eta}{p+\eta}\right)
\theta^{-+}p_{++}\right),\nonumber\\
{\cal Q}^{++}_n=&\oint\dz 
e^{i\left(\frac{n+\eta}{p+\eta}\right)X^+}
\Bigg(p_{--}+\frac{1}{2}i\partial X^+\theta^{++}\nonumber\\
&\hspace{2cm}
+\frac{1}{2}\left(i\partial Z
-\left(\frac{n+\eta}{p+\eta}\right)\theta^{+-}p_{--}\right)\theta^{-+}
+\frac{1}{8}\partial(\theta^{-+}\theta^{++})\theta^{+-}\Bigg),
\nonumber\\
{\cal Q}^{--}_n=&\oint\dz 
e^{i\left(\frac{n-\eta}{p+\eta}\right)X^+}
\Bigg(p_{++}+\frac{1}{2}i\partial X^+\theta^{--}\nonumber\\
&\hspace{2cm}
+\frac{1}{2}\left(i\partial Z^*
-\left(\frac{n-\eta}{p+\eta}\right)\theta^{-+}p_{++}\right)\theta^{+-}
-\frac{1}{8}\partial(\theta^{--}\theta^{+-})\theta^{-+}\Bigg),
\end{align}
on the massless physical states.\cite{HS2} 
These operators include
${\cal P}={\cal P}_p,\ {\cal P}^*={\cal P}^*_{-p},\
{\cal Q}^{\pm\pm}={\cal Q}^{\pm\pm}_{\pm p}$
and generate an affine extension of the supersymmetry algebra (\ref{alg}):
\begin{alignat}{2}
&\left[{\cal J},{\cal P}_n\right]
=\left(\frac{n+\eta}{p+\eta}\right){\cal P}_n,&\qquad
&\left[{\cal J},{\cal P}^*_n\right]
=\left(\frac{n-\eta}{p+\eta}\right){\cal P}^*_n,\nonumber\\
& \left[{\cal P}_n,{\cal P}^*_m\right]
=\left(\frac{n+\eta}{p+\eta}\right){\cal F}\delta_{n+m,0},&\qquad
&\left[{\cal J},{\cal Q}^{\pm\mp}\right]=0,
\nonumber\\
&\left[{\cal J},{\cal Q}^{++}_n\right]
=\left(\frac{n+\eta}{p+\eta}\right){\cal Q}^{++}_n,&\qquad
&\left[{\cal J},{\cal Q}^{--}_n\right]
=\left(\frac{n-\eta}{p+\eta}\right){\cal Q}^{--}_n,\nonumber\\
&\left[{\cal Q}^{-+},{\cal P}_n\right]
=-\left(\frac{n+\eta}{p+\eta}\right){\cal Q}^{++}_n,&\qquad
&\left[{\cal Q}^{+-},{\cal P}^*_n\right]
=-\left(\frac{n-\eta}{p+\eta}\right){\cal Q}^{--}_n,\nonumber\\
& \left\{{\cal Q}^{++}_n,{\cal Q}^{--}_m\right\}
={\cal F}\delta_{n+m,0},&\qquad
& \left\{{\cal Q^{+-}},{\cal Q}^{-+}\right\}={\cal J},\nonumber\\
&\left\{{\cal Q}^{+-},{\cal Q}^{++}_n\right\}
={\cal P}_n,&\qquad
&\left\{{\cal Q}^{-+},{\cal Q}^{--}_n\right\}={\cal P}^*_n.
\end{alignat}
This extended supersymmetry algebra 
provides an enhanced space-time symmetry. 
This symmetry is obtained by taking the Penrose limit of 
${\cal N}=2$ superconformal symmetry, being the isometry of the
$AdS_3\times S^1$ background.\cite{HS2}
It is interesting to trace this limit by studying
hybrid superstrings in $AdS_3\times S^1$. Such an investigation is
currently underway, and it will be reported elsewhere.\cite{K}

The above-mentioned boson-fermion asymmetry 
in the short-string spectrum
does not exist in the RNS formalism, because there are two 
additional physical states in the Ramond sector.\footnote{
The author would like to thank the referee for pointing 
out this fact.} However, this nonequivalence 
of the physical spectrum does not give rise to a tree-level 
inconsistency, because the extra RNS fermions are also 
supersymmetry singlets. It would be interesting to give 
an interpretation of this new spectrum, 
although we have to further check the modular invariance of 
the one-loop partition function for quantum consistency.

It would also be interesting to understand general aspects of 
the holographic duality in plane wave backgrounds.\cite{HS2,KP} 
This will require further investigation. We hope that the manifest 
supersymmetry in the hybrid formalism will shed new light on 
such problems.

\section*{Acknowledgements}
The author would like to thank Y.~Hikida and Y.~Sugawara
for valuable discussions. This work was supported in part by 
Grants-in-Aid for Scientific Research
from the Japan Society for the Promotion of Science (No. 11640276)
and from the Ministry of Education, Culture, Sports, Science 
and Technology of Japan (No.13135213).

\end{document}